\documentclass{iaus}
\usepackage{graphicx}

\title[Group infall of subhaloes at galactic scale] 
{Group infall of substructures on to a \\Milky Way-like dark halo}

\author[Y.-S. Li \& A. Helmi]   
{Yang-Shyang Li \and Amina Helmi}

\affiliation{Kapteyn Astronomical Institute, University of Groningen, \\
P.O. Box 800, 9700 AV Groningen, the Netherlands \\ 
email: {\tt ysleigh@astro.rug.nl; ahelmi@astro.rug.nl} \\[\affilskip]}

\pubyear{2008}
\volume{254}  
\pagerange{1--6}
\jname{The Galaxy Disks in Cosmological Context}
\editors{J. Andersen, J. Bland-Hawthorn \& B. Nordstr\"{o}m, eds.}
\begin{document}

\maketitle

\begin{abstract}
We report the discovery that substructures/subhaloes of a 
galaxy-size halo tend to fall in together in groups  in cosmological 
simulations, something that may explain the oddity of the MW satellite 
distribution.  The original clustering at the time of infall 
is still discernible in the angular momenta of the subhaloes 
even for events which took place up to eight Gyrs ago, $z \sim 1$.  
This phenomenon appears to be rather common since at least $1/3$ 
of the present-day subhaloes have fallen in groups in our simulations.  
Hence, this may well explain the Lynden-Bell \& Lynden-Bell ghostly streams.  
We have also found that the probability of building up a flattened 
distribution similar to the MW satellites is as high as $\sim 80\%$ 
if the MW satellites were from only one group and $\sim 20\%$ when 
five groups are involved.  Therefore, we conclude that the `peculiar' 
distribution of satellites around the MW can be expected with the 
CDM structure formation theory.  This non-random assignment of 
satellites to subhaloes implies an environmental dependence on 
whether these low-mass objects are able to form stars, possibly 
related to the nature of reionization in the early Universe.  
\keywords{Methods: $N$-body simulations, Galaxy: formation, galaxies: dwarf, galaxies: kinematics and dynamics.}
\end{abstract}

\firstsection 
\section{Introduction}
The discrepancy in numbers between the substructures/subhaloes 
resolved in a galaxy-size cold dark matter (CDM) halo and the 
satellites around the Milky Way (MW) has been a long standing 
issue for the `concordance' CDM structure formation theory.  
It implies a non-trivial mapping between the luminous satellites 
and the dark matter subhaloes at the (sub) galactic scales 
\cite[(Zentner \etal\, 2005;]{zentner05} 
\cite[Libeskind \etal\, 2007)]{libeskind07} through the 
astrophysics processes with baryons or the theory needs a 
major modification at a fundamental level (see e.g. 
\cite[Kamionkowski \& Liddle 2000)]{kl00}.

In the past ten years new attention has been drawn to the 
dynamical properties of the MW satellites. Starting with
\cite{lyndenbell95}, the existence of ghostly streams of satellites
(dwarf galaxies and globular clusters) was postulated. These objects
would share similar energies and angular momenta producing a strong
alignment along great circles on the sky.
Recently, the anisotropic distribution of satellites around the 
MW has been argued to be a problem for the CDM theory 
\cite[(Kroupa, Theis \& Boily, 2005;]{kroupa05} 
\cite[Metz, Kroupa \& Jerjen, 2007)]{metz07}.  The 
MW has approximately 20 satellites forming a disk-like structure while the simulated dark matter subhaloes 
usually distribute almost isotropically.  

Here we report our findings of subhaloes falling in groups in 
dark matter simulations and its application on explaining the 
oddities of the dynamical properties of MW satellites, namely, 
the Lynden-Bell \& Lynden-Bell ghostly steams and the great MW 
satellites disk.  Researches in the 
past showed that clusters of galaxies are built of galaxies 
coming in groups \cite[(Knebe \etal\, 2004)]{knebe04}, but it 
was not clear whether a similar picture also applies at the (sub)galactic scale. 
We refer readers to a more detailed  description 
of our analyses and discussions on the group infall and 
its link to the environment in \cite{lihelmi08}.
\section{Substructures in a galaxy-size dark halo}
\subsection{The $N$-body Simulations}
To study the dynamical properties of dark matter subhaloes, 
we have analysed the GAnew series of high resolution simulations 
of a MW-like halo in a full cosmological context ($\Omega_{0}=0.3, 
\Omega_{\Lambda}=0.7$, H$_{0}$=100$h$ km s$^{-1}$ Mpc$^{-1}$ and $h=0.7$).  
The simulations were carried out with GADGET-2 
\cite[(Springel 2005)]{springel05} and a more detailed description 
on the simulation itself are reported in \cite{stoehr06}.  
In the highest mass resolution simulation (GA3new), about 
$10^{7}$ particles within the virial radius resolve the MW-like 
halo at $z = 0$.  These simulations are abundant with self-bound 
substructures ($\sim 4,000$ in GA3new) and the starting redshifts 
of the simulations are as high as $z=37.6$, therefore rendering the 
simulations ideal for studies of subhalo populations and dynamics.

\subsection{Group infall of dark matter subhaloes}

The degree of clustering is quantified by computing the two-point
`angular correlation function', $\omega({\alpha})$, of the
present-day angular momentum of the subhaloes. 

\[\omega(\alpha)=\frac{N(\alpha_{ij}<\alpha)_{\mathrm{simulation}}}{N(\alpha_{ij}<\alpha)_{\mathrm{isotropic}}}-1\]

The angle $\alpha$ is 
the relative orientation of the angular momenta of any two
subhaloes, i.e.  $\cos \alpha_{ij} = \mathbf{L}_{i} \cdot
\mathbf{L}_{j}/(|\mathbf{L}_{i}||\mathbf{L}_{j}|$).  Therefore the
correlation function measures the number of pairs, $N$, with
$\alpha_{ij} < \alpha$ seen in the simulations compared to 
what is expected from an isotropic
distribution with the same number of objects.
An excess of pairs at small angular separations indicate 
small scale clustering in the present-day angular momentum.
This clustering in the angular momentum space is still discernible 
even for some groups accreted about eight Gyrs ago ($z \sim 1$). 

We now focus on the characteristics of groups accreted at
various epochs. To identify groups we link pairs of infalling haloes
whose angular momentum orientations are separated less than ten degrees, i.e., $\alpha <
10^\circ$, and with relative distances $d < 40$ kpc at the time of
accretion. We found that this combination of $\alpha$ and $d$ values
results in a robust set of groups, maximising their extent while
minimising the number of spurious links.

\begin{figure}
\begin{center}
\includegraphics[width=0.95\textwidth]{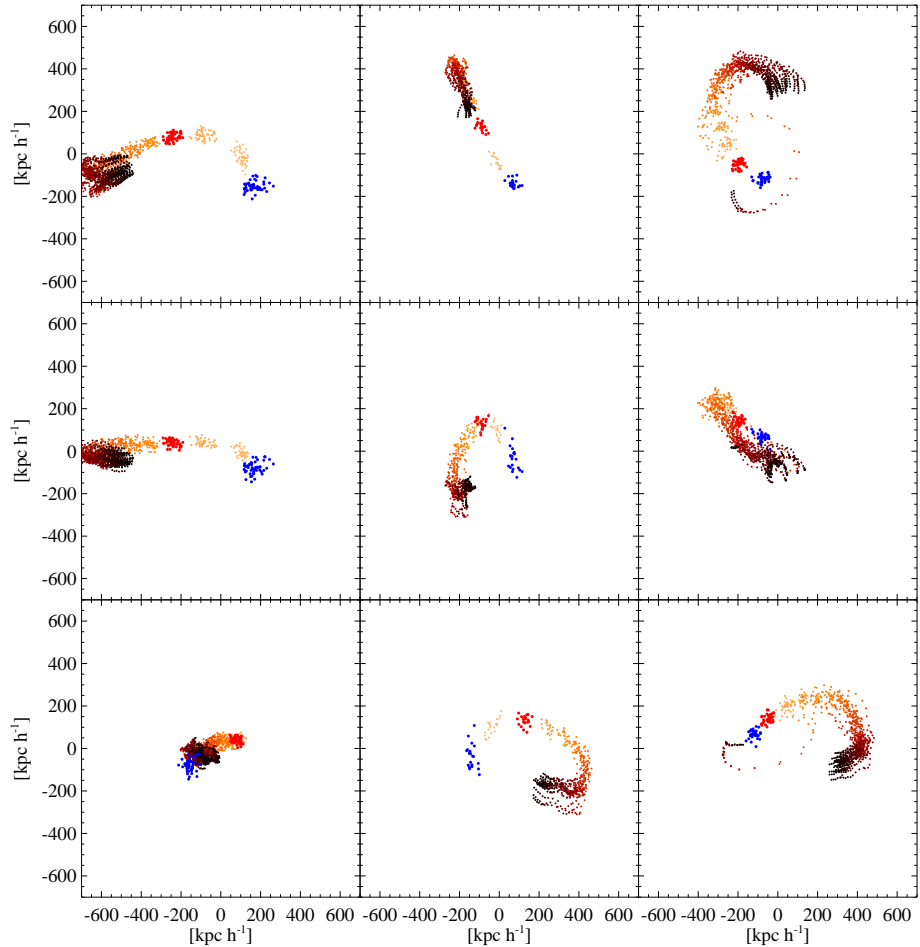} 
\caption{Three examples of the trajectories of groups of
subhaloes accreted at different epochs in the GA3new simulation
reference frame projected on to different planes. 
These are some of the most abundant groups ever accreted.
The colour gradients indicate the arrow of time, from dark at high
redshift to light grey at the present. The positions at accretion
and present time are highlighted with red and blue
respectively.}
\label{fig1}
\end{center}
\end{figure}

We then follow the orbits of the groups identified from redshift $z \sim
4.2$ until present time.  Fig.~\ref{fig1} shows the
trajectories of some of the richest groups of subhaloes, which were
accreted 2.43, 1.65 and 0.84 Gyrs ago respectively.  Each dot
represents the position of a subhalo colour coded from high-redshift
(dark) to the present (light).  The blue symbols correspond to the
present-day positions while those at the time of accretion are shown
in red.  Fig.~\ref{fig1} clearly shows that
groups of subhaloes follow nearly coherent orbits even long before being accreted.

\bigskip

{\underline{\it Mass function of groups}}

We have also looked the mass distribution of groups of subhaloes 
at the time of accretion.  The mass function is dominated 
by low mass groups and can be fitted with a power-law, 
$dN/dlogM \propto M^n$ with $n \sim -0.5 \pm 0.15$.  
Note that this slope is slightly shallower compared to 
the mass function of the full subhalo population 
in a galaxy and a galaxy cluster size dark halo where 
$n = -0.7 - -0.9$ \cite[(e.g., Gao \etal\, 2004)]{gao04}.  At 
the limit of our simulations, we also examined the mass 
spectra within individual groups.  The data suggests the 
mass distribution of subhaloes also follows a power-law 
trend in a group, albeit the signal is much noisier.

\section{On the dynamical peculiarities of the MW satellites}

\subsection{Lynden-Bell \& Lynden-Bell ghostly streams}
Fig.~\ref{fig1} shows the group members following similar 
orbits and retaining the coherence for several Gyrs. 
The clustering of subhaloes in the space is most prominent 
when they accrete on to the MW-like halo.  But later on, the 
spatial coherence of groups will be destroyed due to the gravitational 
interactions with the host and the group members will spread along the orbit.
We note that this characteristic of groups would naturally 
lead to structures sharing common orbital planes, analogues 
to \cite[Lynden-Bell \& Lynden-Bell]{lyndenbell95} 
ghostly streams.  Thus group 
infall provides an explanation clearly different from the 
disruption of a massive progenitor 
\cite[(Lynden-Bell \& Lynden-Bell, 1995)]{lyndenbell95} 
or the tidal-origin scenario \cite[(Kroupa 1997)]{kroupa97}.

\subsection{Great disk of Milky Way satellites}
As stated in the Introduction, the distribution of the MW 
satellites is not isotropic.  \cite{kroupa05} have shown 
that the satellites form a highly flattened `Great 
Satellites Plane' whose \textit{rms} distance to this plane 
is $10-30$ kpc.  These authors conclude 
that this anisotropic structure is inconsistent with that 
of the subhaloes seen in the CDM simulations.  Here we 
reconsider this issue with our high resolution CDM simulations.  In our analyses, we focus 
on the eleven `classical' MW satellites since it is not 
clear whether the newly discovered SDSS satellites confirm the
flattened disk-like structure due to the limited sky coverage.  

\begin{figure}[ht]
\begin{center}
 \includegraphics[width=0.45\textwidth]{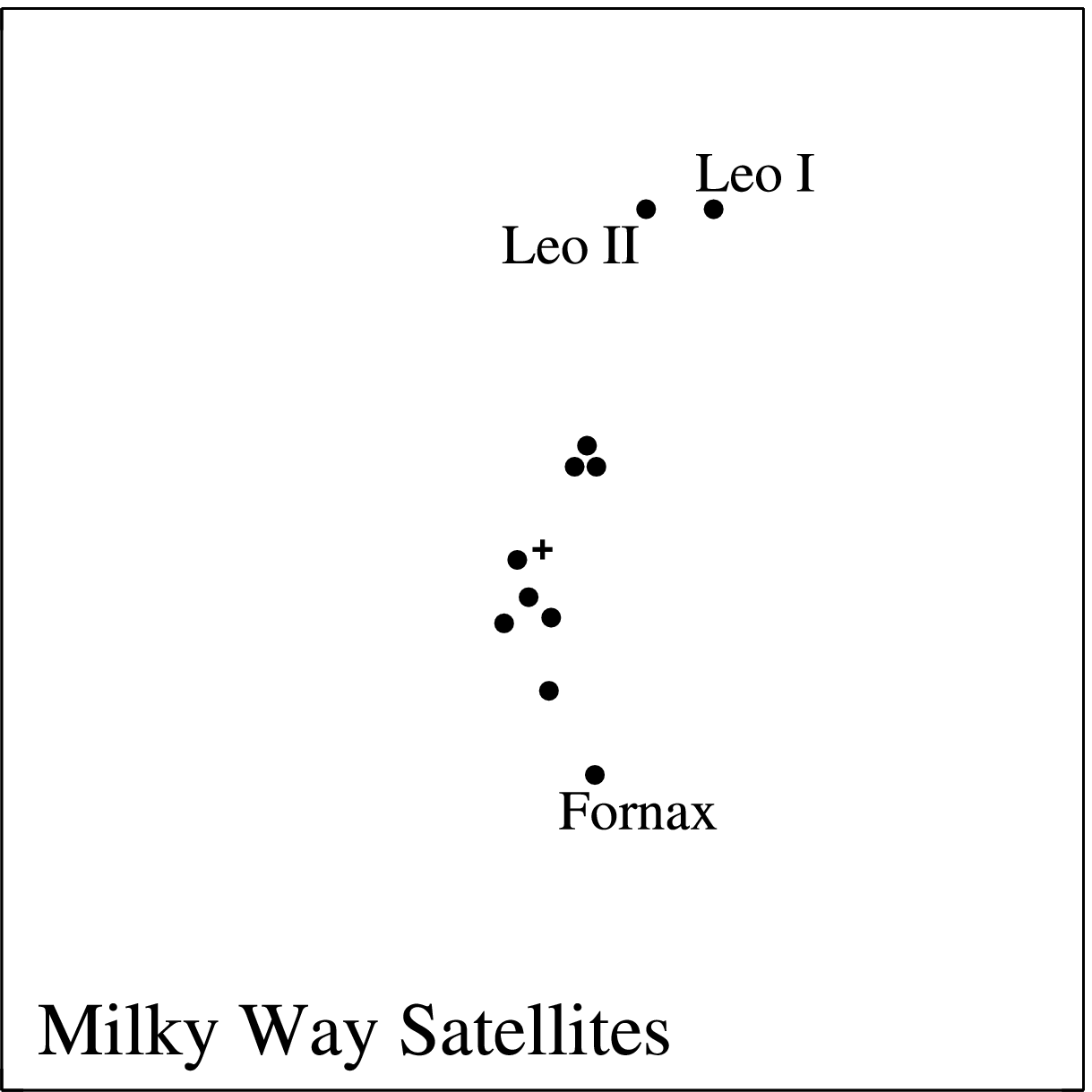} 
 \includegraphics[width=0.45\textwidth]{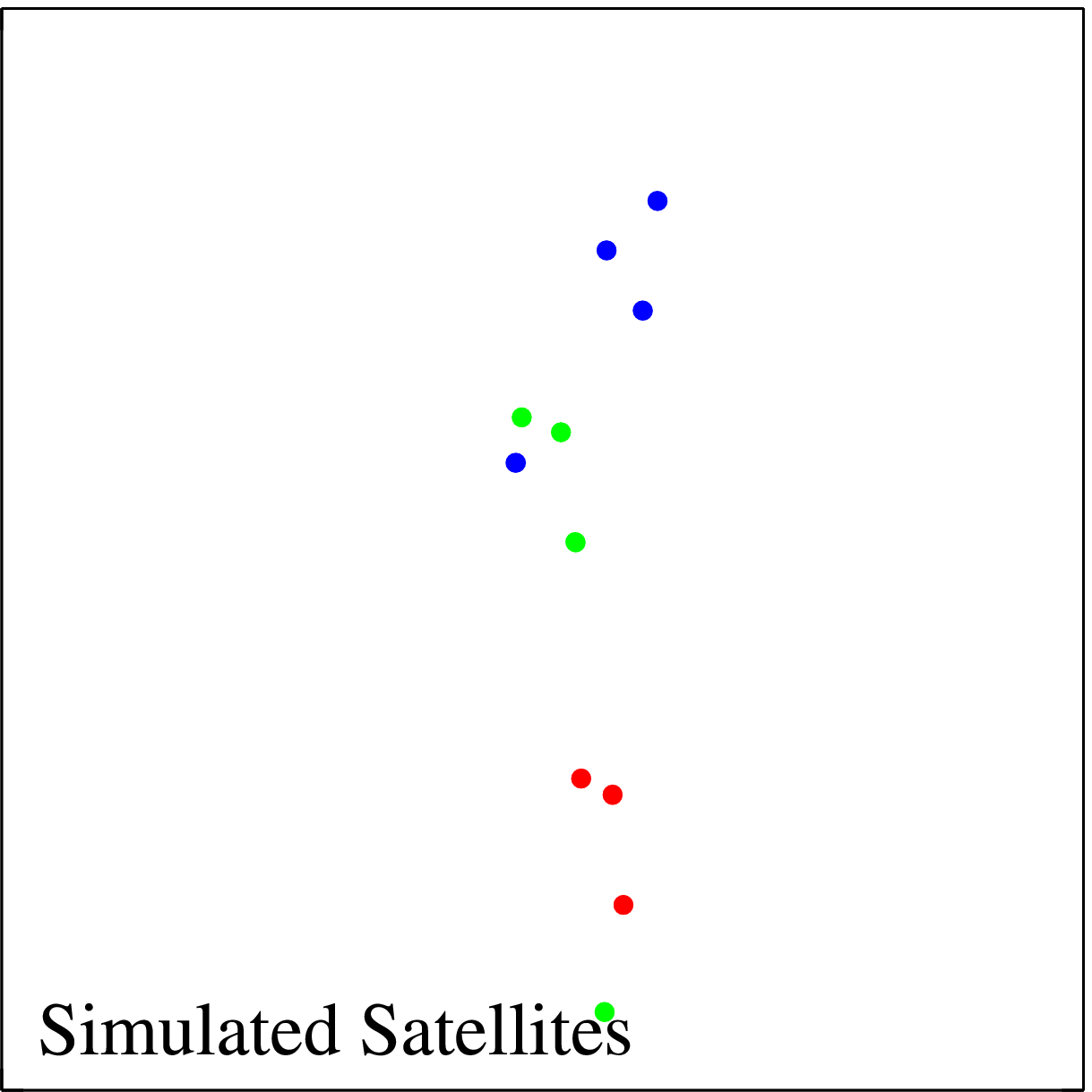} 
 \caption{{\it Left:} Illustration of the distribution of 
the eleven classical Milky Way satellites where the cross 
sign denotes the Galactic centre.  The Galactic disk would 
be horizontal in this orientation.  {\it Right:} An example 
of the distribution of satellites in our simulations which can 
be traced back to come from three different groups as marked by different colours.}
 \label{fig2}
\end{center}
\end{figure}

The great satellite disk of the classical MW satellites has 
an orientation of $72.8\pm 0.7^{\circ}$ with respect to the 
Galactic plane (see Fig.\,\ref{fig2} left panel). 
Here we 
follow the definition by \cite{zentner05} to describe the 
flatness of a satellites disk as $\Delta = D_{rms}/R_{med}$ 
where $D_{rms}$ is the \textit{rms} distance to the 
best-fitted plane, and $R_{med}$ is the median distance of 
objects.  For the MW satellites, the flatness of 
the great disk is $\Delta=0.23 \pm 0.01$ with $D_{rms} \sim 18.5$ and $R_{med} \sim 80$ kpc.  

Of the $3,246$ subhaloes within $300$ kpc from the centre of the MW-like
halo that have survived until the present-day, $898$ subhaloes fell in as
part of a group.  This means about $1/3$ of the surviving subhaloes have joint the MW-like 
halo in the group infall fashion.  $321$ different groups have contributed to the
present-day population of subhaloes, of which the earliest two were
accreted at $z=3.05$.  From now on, surviving subhaloes identified to be part of a
group are referred to as `grouped', while those which are not are
termed `field' subhaloes.

The idea now is to test whether groups of subhaloes would 
form a disk distribution similar to the great MW satellites 
disk.  We have performed two simple tests as follows:

\begin{enumerate}
\item Consider $N_{sub}$ `grouped' subhaloes accreted from only one group and $11-N_{sub}$ from the field. 
\item Consider only the `grouped' subhaloes originated in a few groups.
\end{enumerate}

We generate $10^5$ sets of eleven subhaloes from the present-day 
population which satisfy either of the above mentioned 
conditions and compute the flatness of the best-fit planes.  
The fraction of disks is then the number of sets with flatness 
as low as that of the great satellites disk, i.e., $N(\Delta \le \Delta_{\mathrm{GSD}})$, 
normalised by the total number of realisations.  
The result shows that when considering case (a), the disk fraction
increases from 4.5\% to 73\% as the number of selected subhaloes
$N_{sub}$ increases from $2$ to $11$.  In comparison, 11-randomly selected
subhaloes (within $300$ kpc) gives rise to flattened configurations $\sim
2.2\%$ of the time.  This shows that if the Milky Way satellites fell
in together, it would not be very surprising that they would be in a
planar configuration at the present-day.

When considering only subhaloes originated in groups (case b), 
the fraction of disk-like
configurations obtained in this way can be as high as $\sim 40\%$ when
the subhaloes come from only two groups, and of course reaches 73\%
when they come from just one group.  It is important to note that though 
the disk fraction decreases to $\sim 20\%$ when
selecting from $5$ different groups, it is still much higher than if one
selects $11$ subhaloes randomly.  On the right panel of Fig.\,\ref{fig2}, 
we show an example of the distribution of three groups of subhaloes 
with 3, 4, 4 group members respectively projected along their best-fitted 
plane.  This distribution has $\Delta \sim 0.17$.

Given the large fraction of flattened configurations found in our
simulations, we conclude that the spatial distribution of the $11$ Milky
Way satellites can be reproduced within $\Lambda$CDM. The requirement
is that these satellites fell on to the Galactic halo in groups.
\section{Conclusions and Implications}
We have revisited the issue of the peculiar distribution and
properties of the MW satellites and their link to the 
dark matter subhaloes. In particular, we have focused on the infall 
of substructures on to a Milky-Way like dark matter halo 
in a $\Lambda$CDM cosmogony utilising a series of high-resolution 
dark-matter simulations.  We have found evidence of group infall on to
the MW-like halo, which may explain the ghostly streams proposed by
\cite{lyndenbell95}. 

We have also explored how this planar configuration may be obtained
as a result of the infall of satellites in groups. The observed
correlation in the angular momentum orientation of subhaloes naturally
gives rise to disk-like configurations. For example, we find that if
all subhaloes are accreted from just one group, 
a disk-like distribution is essentially unavoidable ($\sim 80\%$ probability), while for
accretion from just two groups, the likelihood of obtaining a
distribution as planar as observed is 40\%.  Therefore the disky 
configuration of satellites is consistent with CDM if most
satellites have their origin in a few groups.  Note that in our studies, 
we do not need to invoke the baryon-related physics to account for the 
dynamical properties of the MW satellites.  Thus both the `ghostly
streams' and the `planar configuration' are manifestations of the
same phenomenon: the hierarchical growth of structure down to the
smallest galactic scales.

One of the possible implications of the reality of the ghostly streams
is that its member galaxies formed and evolved in a similar
environment before falling into the MW potential. This would
have implications on the (oldest) stellar populations of these
objects, such as for example, sharing a common metallicity floor
\cite[(Helmi \etal\, 2006)]{helmi06}. On the other hand, this 
implies that there should be
groups that have failed to host any luminous satellites. This
would hint at a strong dependence on environment on the ability of a
subhalo to retain gas \cite[(Scannapieco \etal\, 2001)]{scannapieco01}, 
or be shielded from re-ionization by nearby sources 
\cite[(Mashchenko, Carignan \& Bouchard 2004;]{mcb04} 
\cite[Weinmann \etal\, 2007)]{weinmann07}.

Recent proper motion measurements of the Large and Small Magellanic
clouds by \cite{smc-mu}, as well as the simulations by \cite{bc05}
suggest that these systems may have become bound to each other only
recently.  This would be fairly plausible in the context of our
results.  The Clouds may well have been part of a recently accreted
group and it may not
even be necessary for them to ever have been a binary system. 

\section*{Acknowledgements}
YSL thanks the organising committee for giving the opportunity 
to contribute this talk and the travel grant supported by IAU.
This work has been supported by a VIDI grant from the Netherlands 
Foundation for Scientific Research (NWO).

\end{document}